\documentclass[]{spie}  %>>> use for US letter paper
\usepackage[]{graphicx}

\def\deg{\hbox{$^\circ$}}

\def\lae{\mathrel{\raise .4ex\hbox{\rlap{$<$}\lower 1.2ex\hbox{$\sim$}}}}
\def\gae{\mathrel{\raise .4ex\hbox{\rlap{$>$}\lower 1.2ex\hbox{$\sim$}}}}

\title{The Use of Laterally Graded Multilayer Mirrors for Soft X-ray Polarimetry} 

\author{Herman L.\ Marshall\supit{a},
Norbert S.\ Schulz\supit{a},
David L. Windt\supit{b},
Eric M.\ Gullikson\supit{c}
Eric Blake\supit{d},
Dan Getty\supit{a}, 
Zane McInturff\supit{e},
\skiplinehalf
\supit{a}MIT Kavli Institute, Cambridge, MA, USA 02139\\
\supit{b}Reflective X-ray Optics, 1361 Amsterdam Ave, Suite 3B, New York, NY, USA 10027\\
\supit{c}Lawrence Berkeley National Lab, 1 Cyclotron Rd., Bldg. 2R0400, Berkeley, CA, USA 94720\\
\supit{d}University of Massachusetts, Lowell, MA, USA 01854\\
\supit{e}University of Wisconsin, Madison, WI, USA 53706
}

\authorinfo{Further author information: (Send correspondence to H.L.M.)\\H.L.M.: hermanm@space.mit.edu,
Telephone: 1 617 253 8573} 

\begin{document} 
  \maketitle 

%%%%%%%%%%%%%%%%%%%%%%%%%%%%%%%%%%%%%%%%%%%%%%%%%%%%%%%%%%%%% 
\begin{abstract}
We present continued development of laterally graded multilayer mirrors (LGMLs) for a telescope design
capable of measuring linear X-ray polarization over a broad spectral band.
The multilayer-coated mirrors are used as Bragg reflectors at the Brewster angle.
By matching to the dispersion of a spectrometer, one may take advantage of high multilayer
reflectivities and achieve modulation factors over 50\% over the entire 0.2-0.8 keV band.
In Phase II of the polarimetry beam-line development, we demonstrated that the system
provides 100\% polarized X-rays at 0.525 keV (Marshall et al. 2013).
Here, we present results from phase III of our development, where a LGML
is used at the source and laterally manipulated in order to select and polarize
X-rays from emission lines for a variety of source anodes.
The beam-line will then provide the capability to test polarimeter components
across the 0.15-0.70 keV band.
We also present plans for a suborbital rocket experiment designed to detect a polarization
level of better than 10\% for an active galactic nucleus.

\end{abstract}

%>>>> Include a list of keywords after the abstract 

\keywords{X-ray, polarimeter, astronomy, multilayer, mirror, grating}

%%%%%%%%%%%%%%%%%%%%%%%%%%%%%%%%%%%%%%%%%%%%%%%%%%%%%%%%%%%%%
\section{INTRODUCTION}
\label{sec:intro}  % \label{} allows reference to this section

We continue our investigation and laboratory work to develop a soft
X-ray polarimeter based on Bragg reflection from multilayer-coated optics.
Marshall (2007\cite{2007SPIE.6688E..31M})
described a method using transmission gratings
to disperse the incoming X-rays so that the dispersion
is matched to laterally graded multilayer (ML) coated reflectors.  
An extension of this approach was suggested by
Marshall (2008\cite{2008SPIE.7011E..63M})
that can be used with larger missions
such as the AXSIO or AEGIS.
Some potential scientific investigations that
would be possible with a soft X-ray polarimeter
were described earlier and include testing the
synchrotron nature of quasar jet emission and
models of neutron star atmospheres\cite{plexas,2010SPIE.7732E..12M}.

The laboratory work was initiated in order to
demonstrate polarization measurements using gratings and laterally
graded multilayer coated mirrors (LGMLs) for eventual use in a flight design.
Here, we
describe the first results from Phase III of our development work
along with reflectivity data for a new LGML.
A description of a design for a suborbital rocket
flight is given in \S\ref{sec:instr}, comparable to that described in Paper I.
The experiment's  minimum detectable polarization (MDP) is
expected to be about 10\% when observing a bright blazar such
as Mk 421.

\section{The MIT Polarimetry Beamline}

We recommissioned the
X-ray grating evaluation facility (X-GEF), a 17 m beamline that was developed
for testing transmission gratings fabricated at MIT for
the {\it Chandra} project\cite{1994SPIE.2280..257D}.
The project development is proceeding in four distinct phases, of which
two have been completed.  In Phase I, we set up the polarized X-ray
source at one energy (0.525 keV) and aligned it so that the beam was
uniform at the detector and its intensity did not vary
significantly with rotation angle.  In Phase II, we added a ML coated mirror
to the detector end of the system, reoriented the
detector to face 90$\deg$ to the beamline, and demonstrated that
the source produced nearly 100\% polarized X-rays.  Results from
phases I and II were reported by
Marshall et al.\ (2013\cite{2013SPIE.8861E..1DM}, hereafter Paper I).

We are now in Phase III, where we replaced the source ML
coated mirror with a LGML.  In Phase IV, we will insert a diffraction grating
in order to disperse the polarized input onto an LGML in the detector
chamber.  In addition, we will test LGMLs with
improved reflectivities and with larger ML coating periods in order to
demonstrate that they can be used in a flight system, such as the
one described in \S\ref{sec:instr}.

\subsection{Polarimetry Beamline Phase III}

With MKI technology development funding, we 
adapted the source to produce polarized X-rays at the O-K$\alpha$
line (0.525 keV)\cite{2010SPIE.7732E..108M}.
A five-way chamber was added to house the polarized source multilayer (ML) mirror.
The mirror and a twin were provided by Reflective X-ray Optics (RXO), with a coating consisting
of 200 layers of 5.04\AA\ of W alternating with 11.76 \AA\ of B$_4$C.
The wavelength, $\lambda$, of the Bragg peak for a periodic ML coating is given by
$\lambda = 2 d \sin \theta$, where
$d = 16.80$\AA\ is the average layer thickness, and $\theta = 45\deg$ is the
graze angle (measured from the surface).
The detector is a front-side illuminated CCD.

\begin{figure}
   \centering
   \includegraphics[width=15cm]{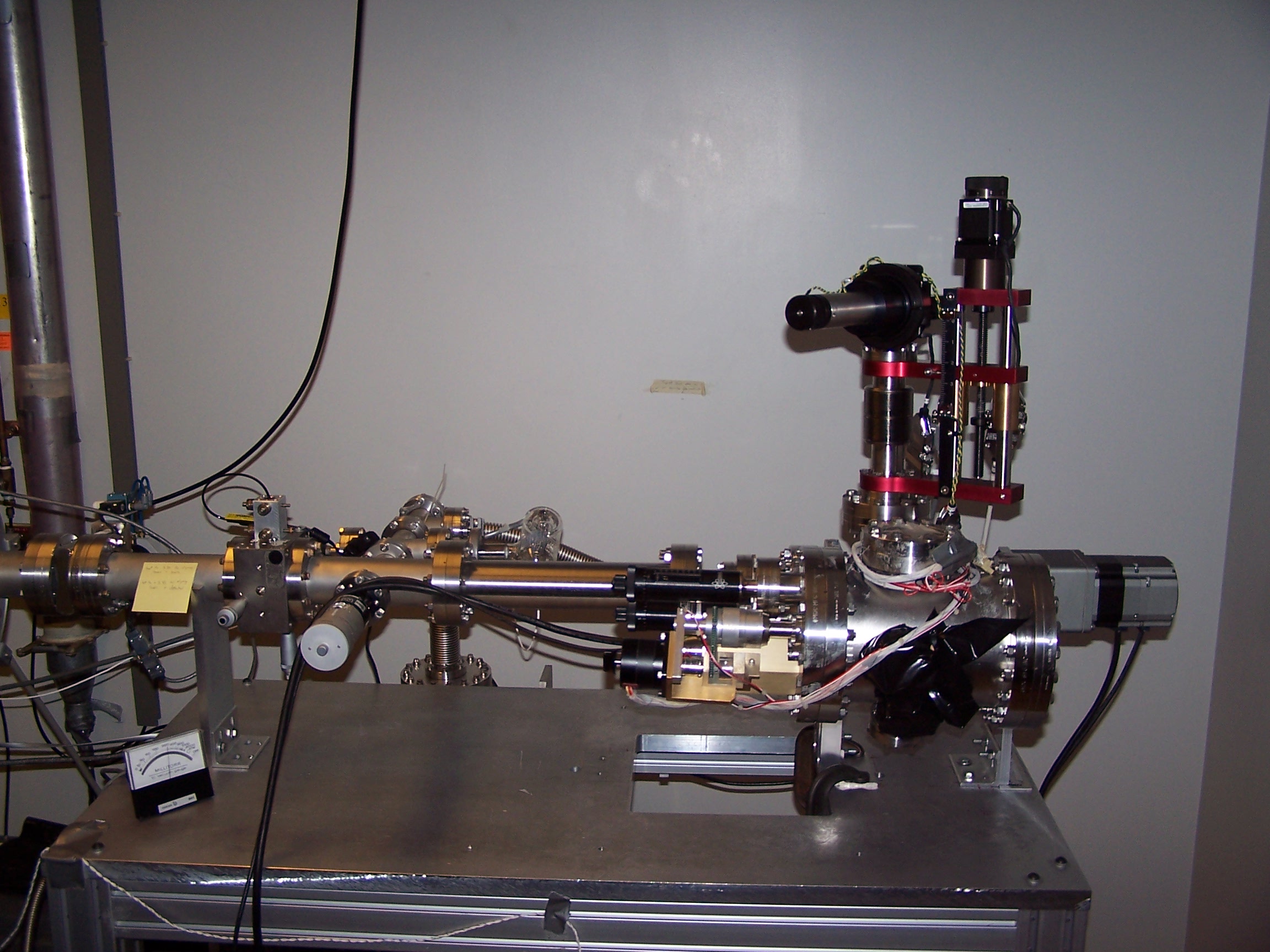}
 \caption{
 The polarizing X-ray source, as configured for development Phase III.
 The X-ray source is horizontal and the linear manipulator (oriented vertically here)
 is used
 to adjust the position of the LGML in the source chamber.  The motor,
 at right, rotates the entire assembly about the beamline axis.
  }
\label{fig:source1}
\end{figure}

\begin{figure}
   \centering
   \includegraphics[width=6.5cm]{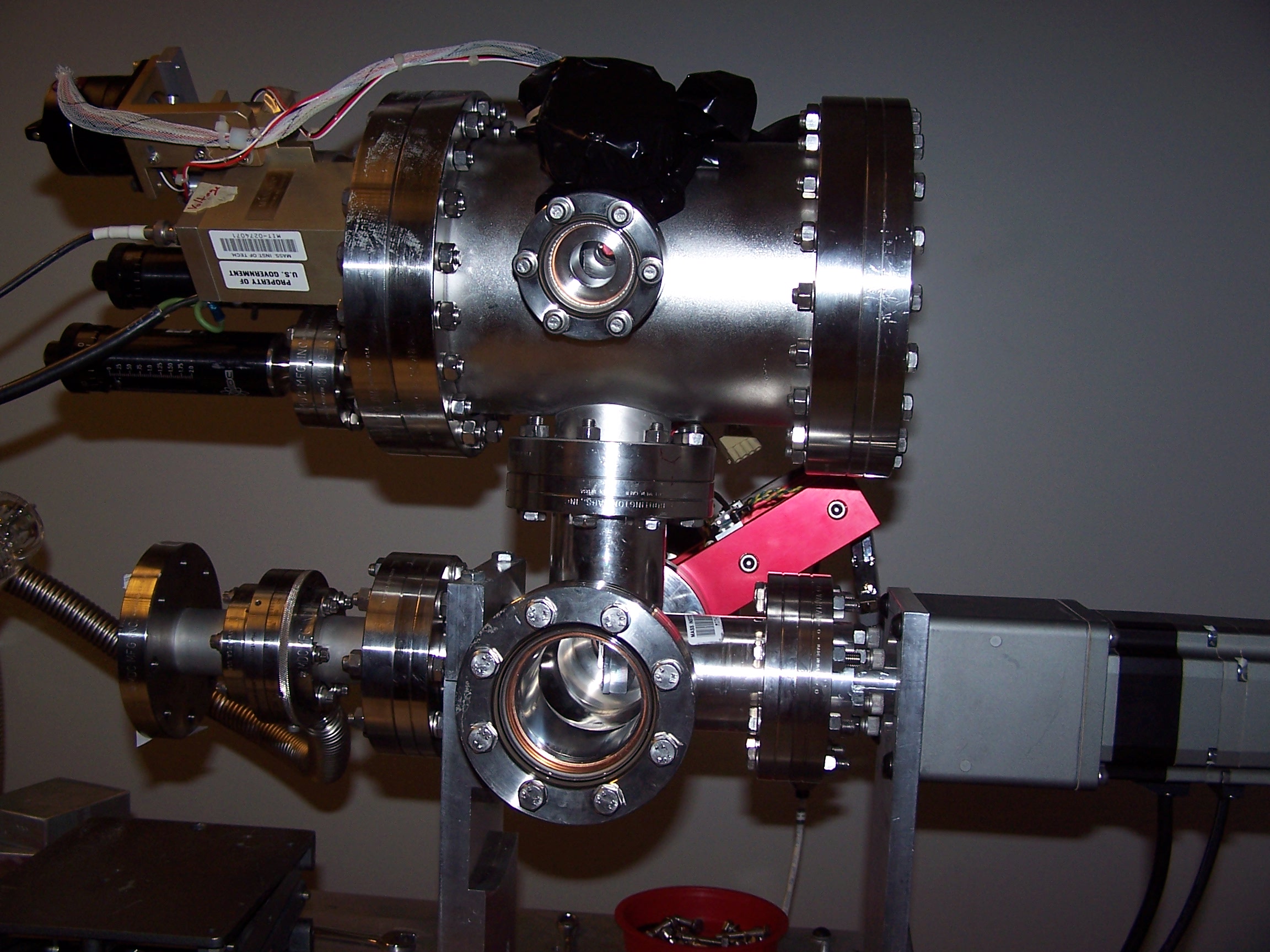}
   \includegraphics[width=3.5cm]{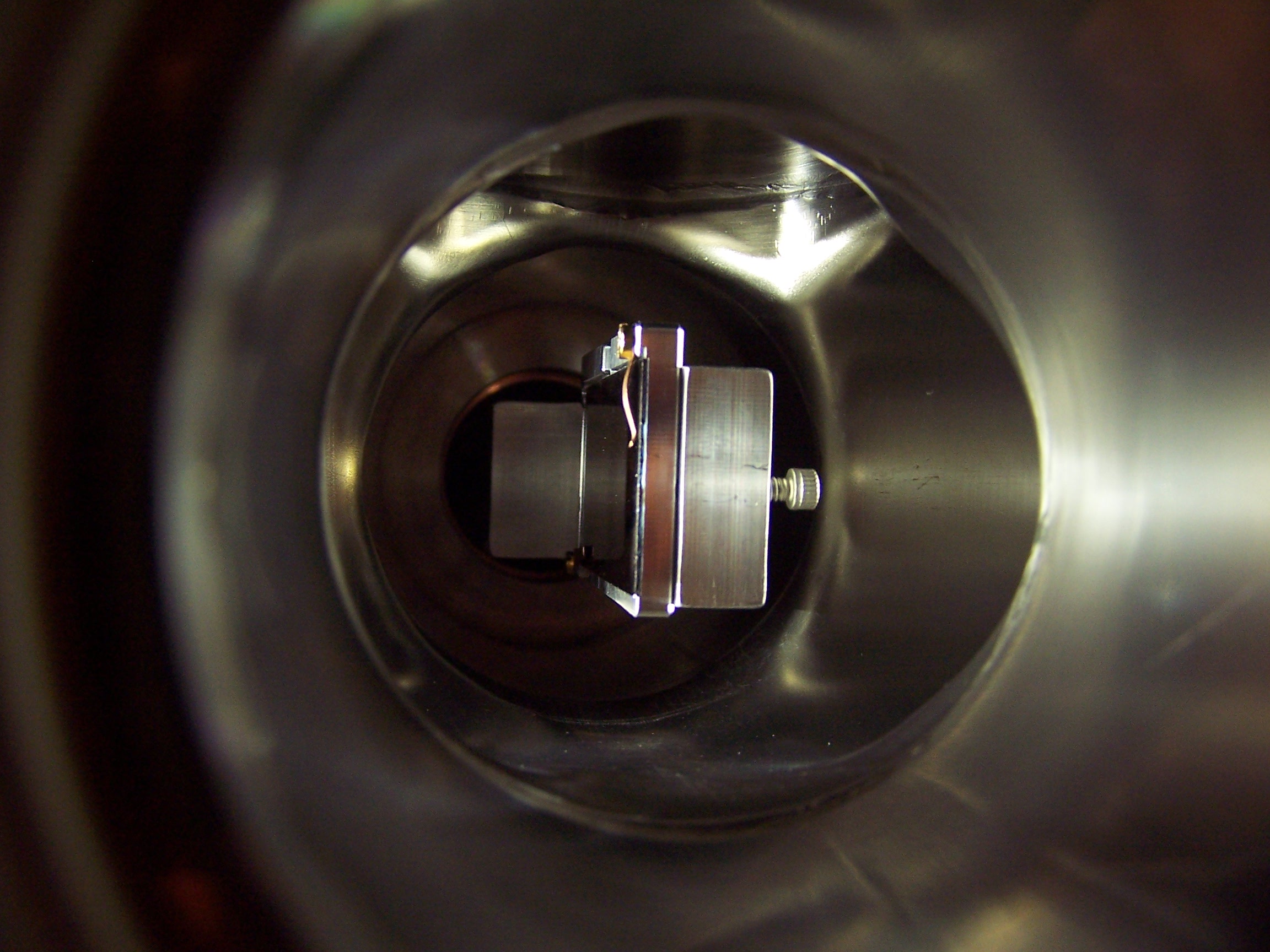}
   \includegraphics[width=6.5cm]{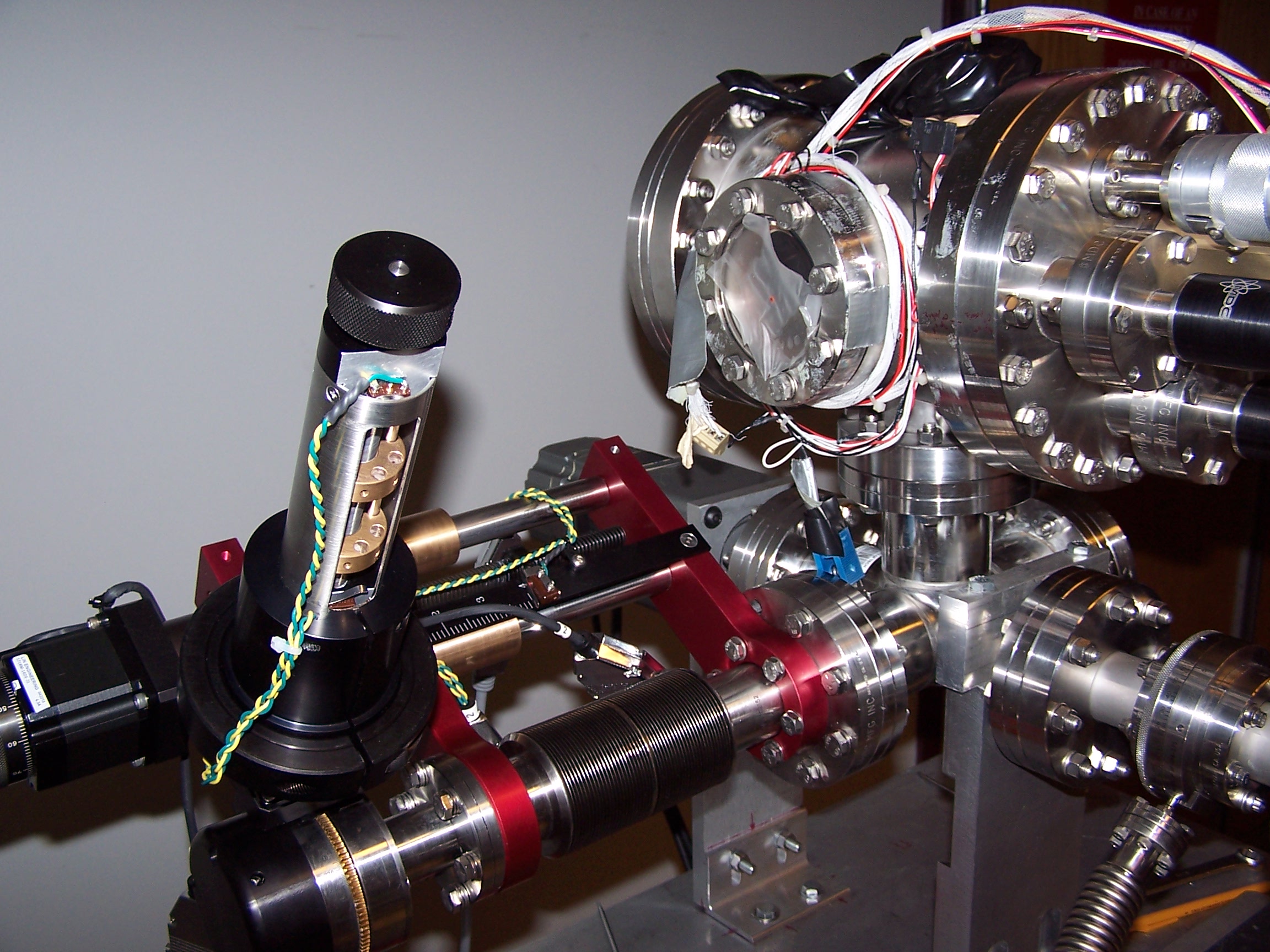}
 \caption{
 Detail of the polarizing X-ray source.
 {\it Left:} The motor was rotated to place the X-ray source above the 
  chamber containing the LGML.  A section of source pipe has been removed for the alignment
  procedure and the LGML is oriented in order to reflect a laser beam from
  the detector chamber back through the collimation aperture.
   {\it Middle:}  Closeup of the LGML that is mounted in the source chamber.
   A linear manipulator moves the mirror along its length in order to reflect
   and polarize a specific energy, usually an emission line of the X-ray source.
   {\it Right:} Manipulator side of the polarizing X-ray source.  A linear
   manipulator (with red anodized base plates) has a 100 mm
   range moving the rotary manipulator.  The LGML is mounted to the shaft
   of the rotary manipulator.
  }
\label{fig:source2}
\end{figure}

With funding from a MIT Kavli Investment grant,
we started Phase III, where the goal was to develop, test, and install a laterally
graded ML coated mirror (LGML) in the source mirror chamber.
A pair of LGMLs were fabricated by RXO, consisting of 80 bilayers of W and
B$_4$C on highly polished Si wafers.
The multilayer period was varied linearly along the 47 mm length of the substrate in order
to reflect and polarize X-rays from 17\AA\ to 73\AA\ (170 to 730 eV).
Fig.~\ref{fig:source1} shows a picture of the source hardware and
Fig.~\ref{fig:source2} shows details.

The W/B$_4$C LGML is now mounted to the shaft of the motor-controlled
rotational manipulator so that the axis of
the shaft is centered on the long axis of the LGML's surface.  The shaft's rotation angle
is controllable to 0.01$\deg$.  It is mounted on a linear bellows with
a 100 mm travel that is computer-controlled to an accuracy of 0.01 mm.

The LGML mirror was aligned by setting up a laser in the detector chamber
about 16 m from the X-ray source and pointed through the square
collimating aperture to the polarizing mirror, thus defining the beamline
optical axis.  This procedure was developed and implemented successfully
in phase I (see Paper I).
The mirror readily reflects
laser light, so it was rotated to be perpendicular to the laser beam,
in order to reflect the laser beam back to the collimating aperture.
The mirror holder has two adjustment screws
(one can be seen in Fig.~\ref{fig:source2}, center), which were adjusted
to ensure that the reflected laser beam was centered on the
collimating aperture, even as the mirror chamber was rotated.  In
this manner, the mirror's axis was aligned to the beamline axis
to within 1 cm along an 8.5 m length, for an accuracy of
about $\pm 0.001$ radians.  Being aligned while oriented
90$\deg$ to the beamline, the LGML was then rotated precisely
45$\deg$ for operation, using the rotational manipulator.

 \begin{figure}
  \centering
   \includegraphics[width=8cm]{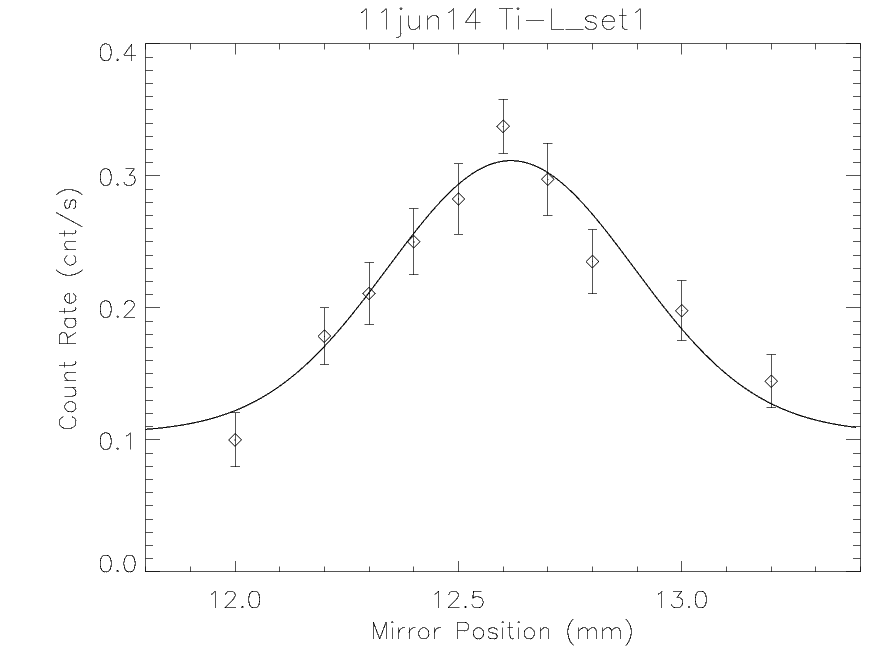}
   \includegraphics[width=8cm]{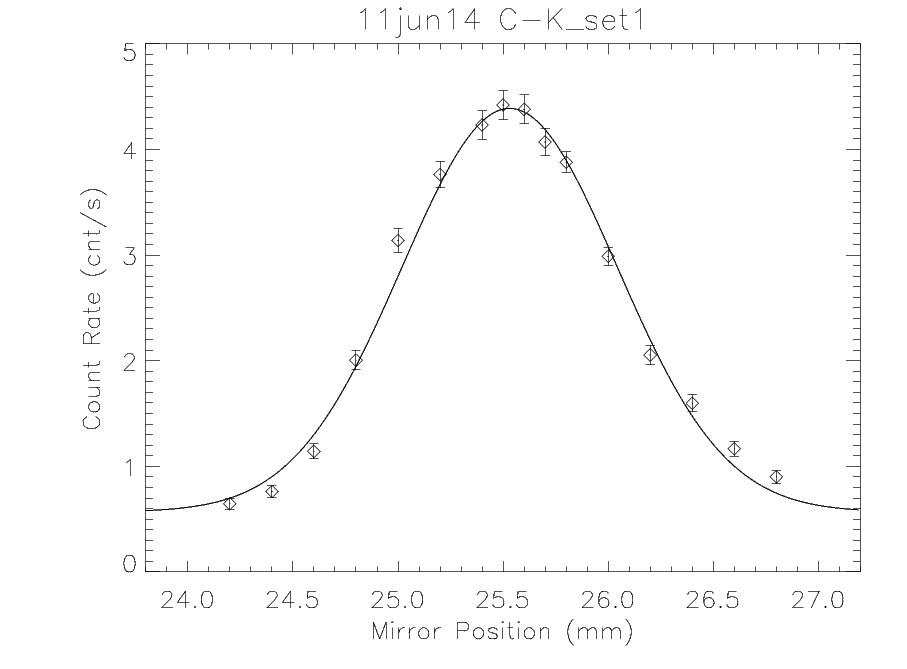}
 \caption{
 Results from Phase III operations of the polarimetry beamline, where the CCD
  receives the polarized beam directly.
 {\it Left:} The CCD count rate as a function of position along the LGML
 when using the Ti source anode.
 The peak due to the Ti-L line complex
 was expected at $x = 12.38$ mm, which is consistent with
 LGML mounting uncertainties.
 {\it Right:} The same as the left panel except that the C anode was used.
 The peak is due to the C-K line.
}
\label{fig:phase3}
\end{figure}

The $x = 0$ location was defined as the edge of the LGML, and $x$
increases along its length.
Operationally, this position was placed at the center of the beamline
to set the linear manipulator's zero point.  The alignment laser was used,
providing a zero point that matched $x = 0$ to within about 0.5 mm.
The location of the Bragg peak for wavelength $\lambda$ was computed
from ALS measurements to be $\lambda = a + b x$, where $a =$
12.076 \AA\ and $b =$ 1.2398 \AA\ mm$^{-1}$.  Inverting, the
predicted locations of the Ti-L$\alpha$ and C-K$\alpha$ lines were
expected to be found at 12.375 mm and 26.36 mm, respectively.
Scans of the linear manipulator are shown in Fig.~\ref{fig:phase3}.

Given that the CCD is front-side illuminated, the C-K line is gratifyingly
bright and the events were readily distinguished from background
events using pulse height discrimination.  Optical light from the
source filament was found to be a major contributor to the background
and was reduced by placing an aperture stop made from aluminum
foil between the source LGML and the X-ray source itself.

The widths of the response curves in Fig.~\ref{fig:phase3} are a the
result of a combination of the breadth of the Bragg reflectivity curve
of the LGML and of the intrinsic width of the C-K line or the Ti-L complex.
From ALS measurements,
the FWHM of the Bragg peak is about 1.3\% of the peak wavelength,
corresponding to a FWHM of 0.29 mm (0.47 mm) for Ti-L (C-K). 
The observed FWHM values are 0.65 mm and 1.20 mm for Ti-L and C-K,
respectively, leaving a contribution of 0.58 mm and 1.10 mm due to the
intrinsic widths of the emission features.  Translated to wavelength
and energy units gives 0.72 \AA\ and 1.37 \AA\ or 11.9 eV and 8.5 eV
for the FWHMs of the Ti-L and C-K emission.

\subsection{Polarimetry Beamline Phase IV}

\begin{figure}
  \centering
   \includegraphics[width=14cm]{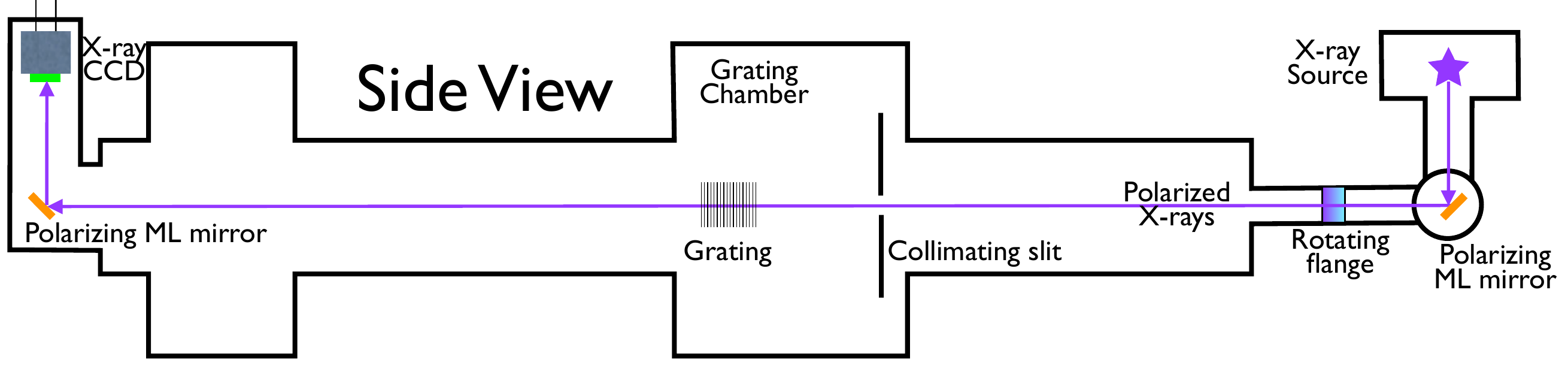}
   \includegraphics[width=2.5cm]{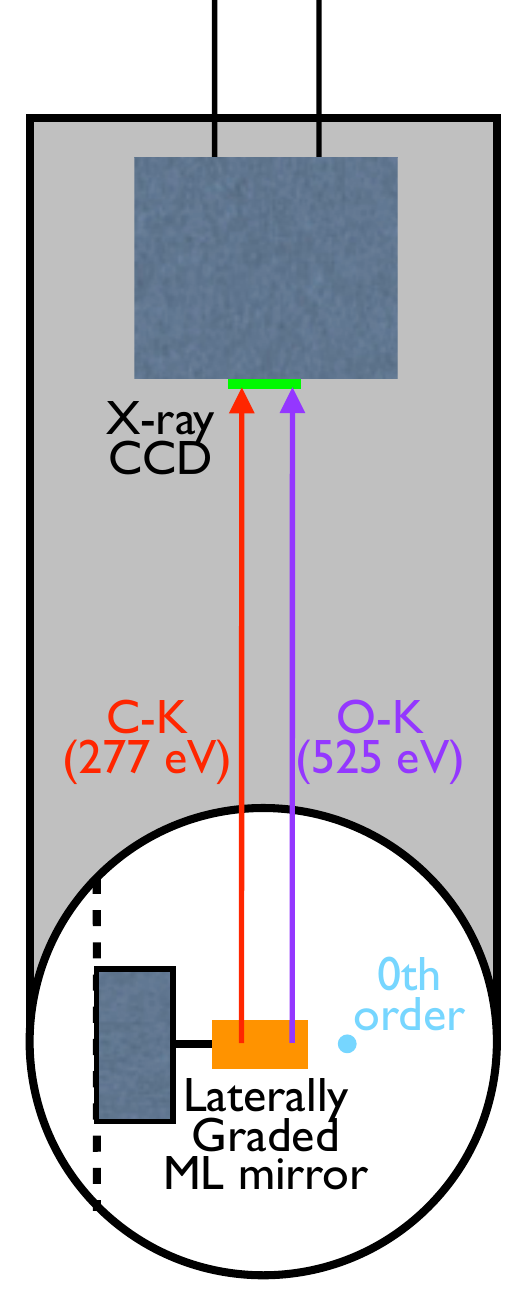}
 \caption{
 Schematic of the polarimetry beamline in Phase IV.  The configuration is the same
 as in Phase II (Paper I) except that the polarizing ML mirrors at
 both ends of the system are replaced with LGMLs and there is a grating
 mounted in the grating chamber to disperse X-rays to specific locations on
 the detector LGML.
 {\it Left:} Side view, showing the location of the grating.
 {\it Right:} Beam view, showing that the LGML will reflect different input energies
 from different locations on its surface.
}
\label{fig:schematic4}
\end{figure}

With funding from the NASA Astrophysics Research and Analysis (APRA) program,
we will soon move to Phase IV, where the goals are 1) to improve the reflectivities of LGMLs
by trying new material combinations and 2) show that a grating-LGML combination
can measure polarization over wide range of energies, thus prototyping a design
that could be used for a flight system.  See Fig.~\ref{fig:schematic4} for a
schematic of the Phase IV configuration.

\begin{figure}
   \centering
   \includegraphics[width=8cm]{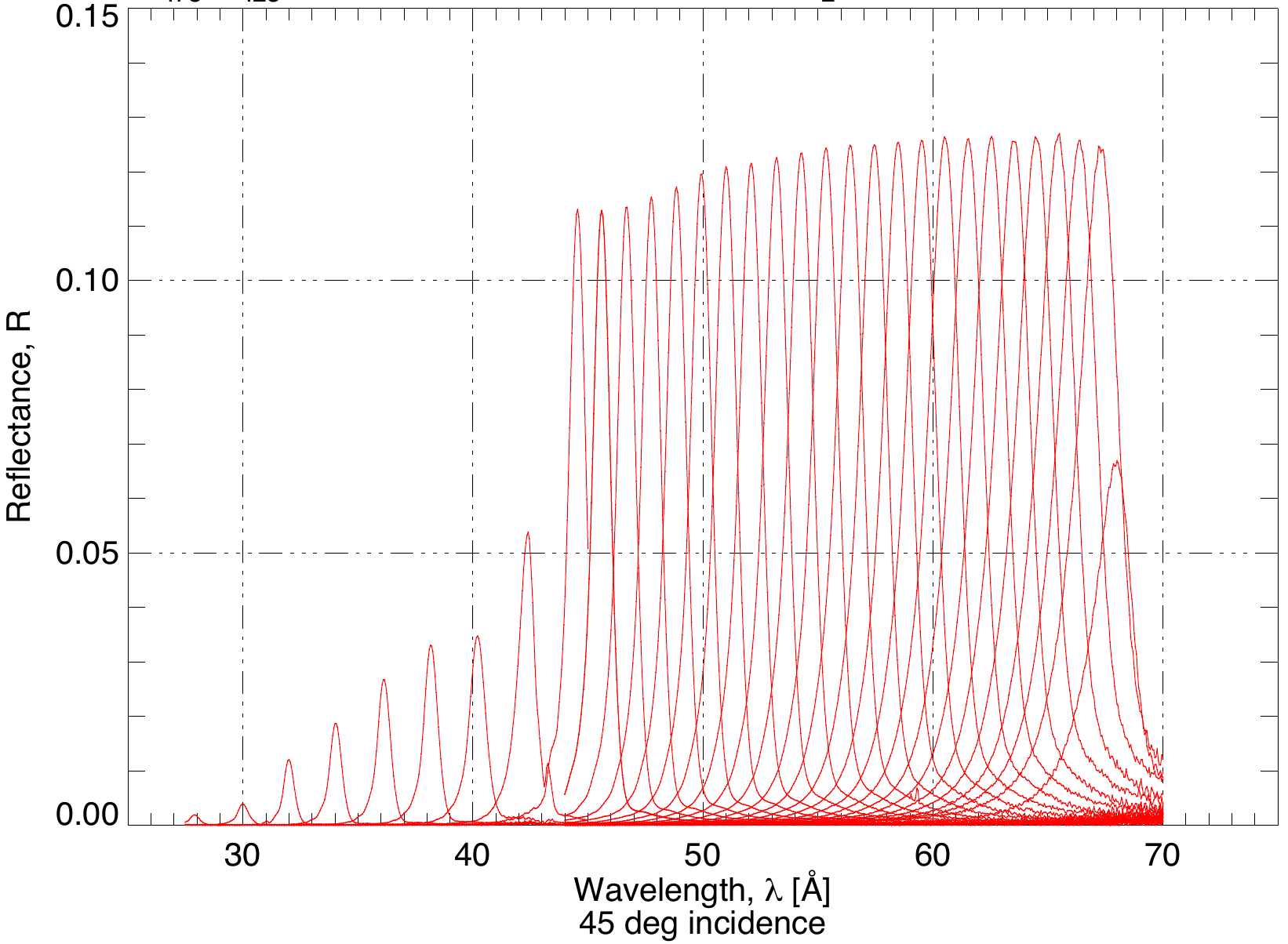}
   \includegraphics[width=8cm]{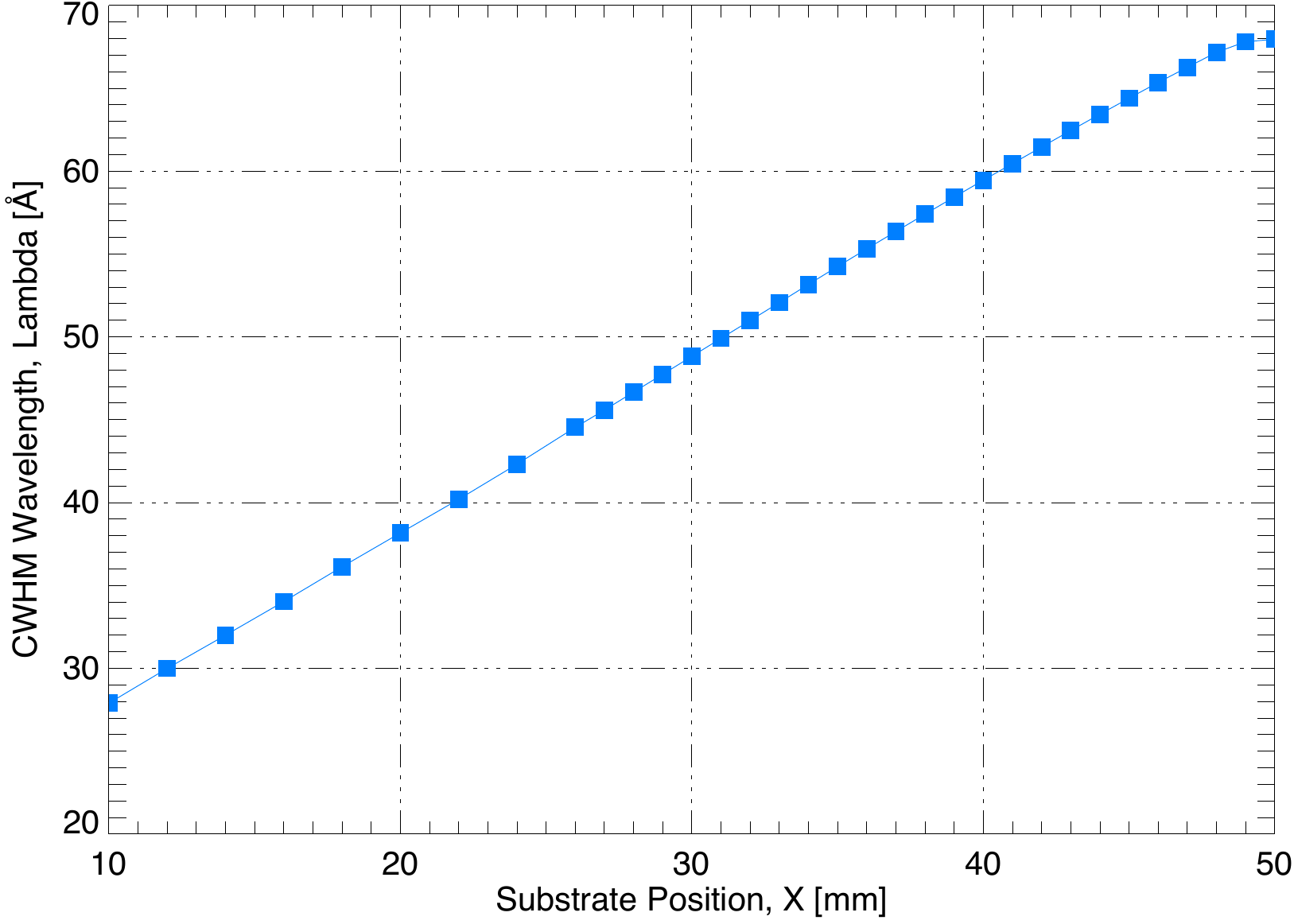}
 \caption{
Multilayer reflectivities for a laterally graded ML (LGML) made by
  Reflective X-ray Optics (RXO) using layers of C interspersed with layers
  of an alloy of Cr and Co.
  {\it Left:} Reflectivity measurements using the ALS.
  For one measurement run, the reflectivity was sampled at 2 mm spacing, starting
  below 44 \AA.  In a second run (above 44 \AA), the samples were at 1 mm spacing.  The ALS
  beam consisted of about 70\% s-polarization and data were taken at a graze
  angle of 45$\deg$.
  {\it Right:} Bragg peak as a function of position along the LGML.
}
\label{fig:ccr}
\end{figure}

\begin{figure}
   \centering
   \includegraphics[width=7cm]{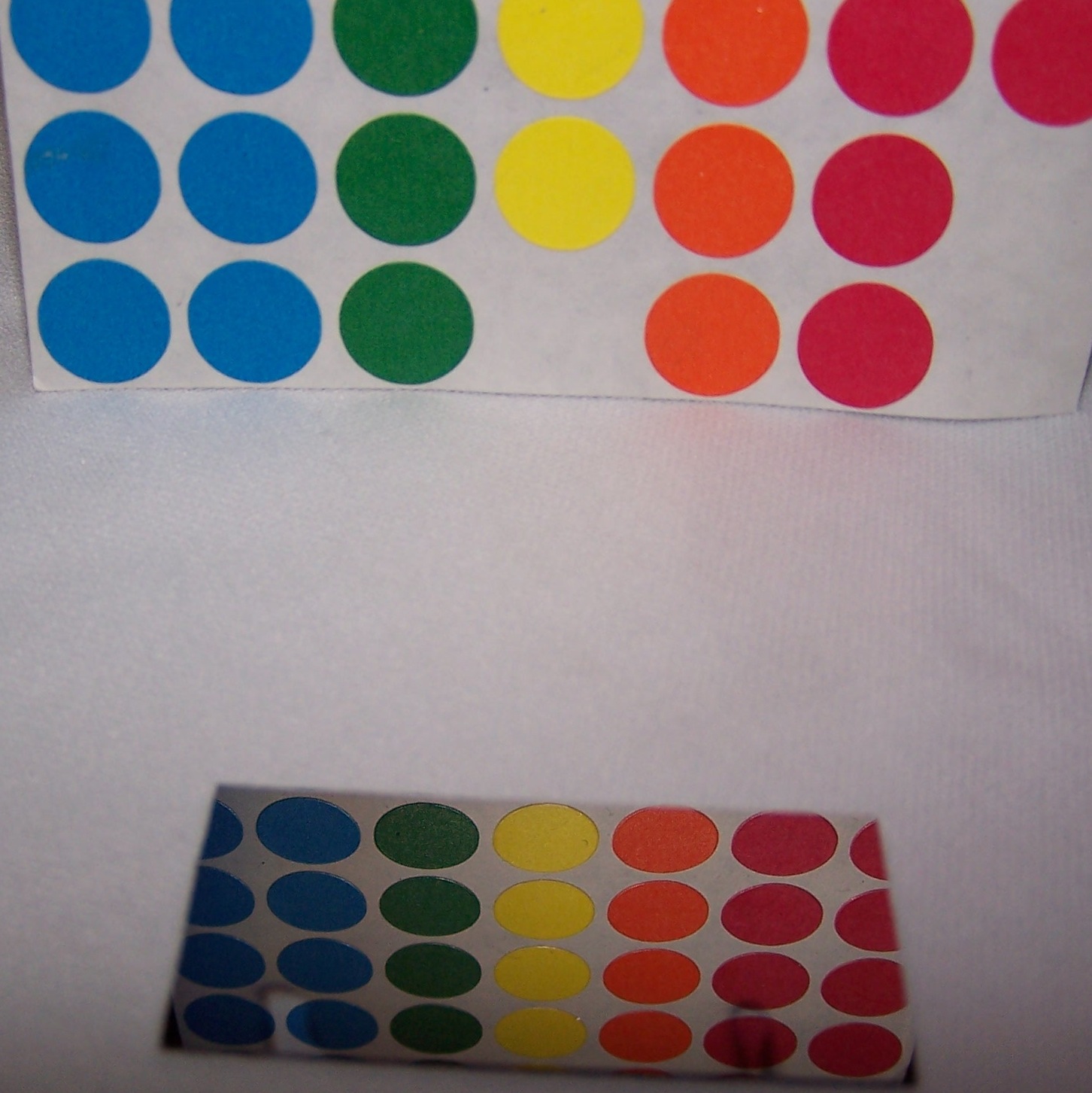}
 \caption{
 A laterally graded ML coated mirror (LGML) from RXO
  using layers of C interspersed with layers of an alloy of Cr and Co.
  It is about 47 mm long, 23 mm wide, and 0.5 mm thick.
  The ML period increases from left to right.
   }
\label{fig:lgml}
\end{figure}

We have begun work to improve LGML reflectivities using different
ML compositions for specific wavelength regions.  Figure~\ref{fig:ccr}
shows results from ALS testing of the first few single-period MLs involving
C and Cr or a CoCr alloy, shown in Fig.~\ref{fig:lgml}.
The C/CoCr reflectivities to s-polarization approach 20\% in the 45-65 \AA\ range.
These reflectivites are substantially better than the 2-5\% values currently
available in the W/B$_4$C LGMLs (Paper I).
ML coating process adjustments will be varied in order to improve the
reflectivities below 45 \AA.

The main advance in Phase IV will be to use gratings to disperse the X-rays
to the LGML in the detector chamber.  The dispersion of the grating is given by the
grating equation: $m \lambda = P \sin \phi$, where $P$ is the grating period, $m$
is the grating order of interest (which we take to be $+1$) and $\phi$ is the
dispersion angle.  Defining $y$ to be the horizontal direction in Fig.~\ref{fig:schematic4}(right)
and defining $y = 0$ to be where the 0th order lands at the plane of the LGML,
then we match the LGML's Bragg peak to the grating dispersion by setting
$P \sin \phi = P y/D = 2 d(y) \sin \theta = \surd 2 ~ d(y)$, where $D$ is
the distance from the grating to the LGML, giving $d = P y / (D \surd 2)$ as
the multilayer period.
As long as $d(y)$ is linear, the LGML can be placed at distance $D$
from the grating to reflect X-rays of arbitrary wavelengths, within the physical limitation
of the LGML.  The current LGMLs have gradients d$d/$d$y = 0.877 $\AA/mm,
giving d$\lambda/$d$y = 1.240 $\AA/mm, which is
matchable by gratings made for the {\it Chandra} Low Energy Transmission
Grating (LETG) Spectrometer \cite{1997SPIE.3113..172P} with $P = 9912$\AA\ for $D = 8.0$ m.
We have four LETG facets
on loan from MPE (courtesy P. Predehl) that
will be mounted in the grating chamber for this purpose.
When completed, the Phase IV configuration emulates a flight system from the
point of the beam line from the rotating flange down to the detector.  By
varying the linear position of the LGML mounted in the source chamber,
we will be able to generate 100\% polarized X-rays over the
17-73 \AA\ (170 to 730 eV) range and rotate the polarization angle
to confirm modulation curves such as obtained in Phase II (see Paper I).

\section{Moving the Polarimetry Beamline}

\begin{figure}
    \centering
   \includegraphics[height=11cm]{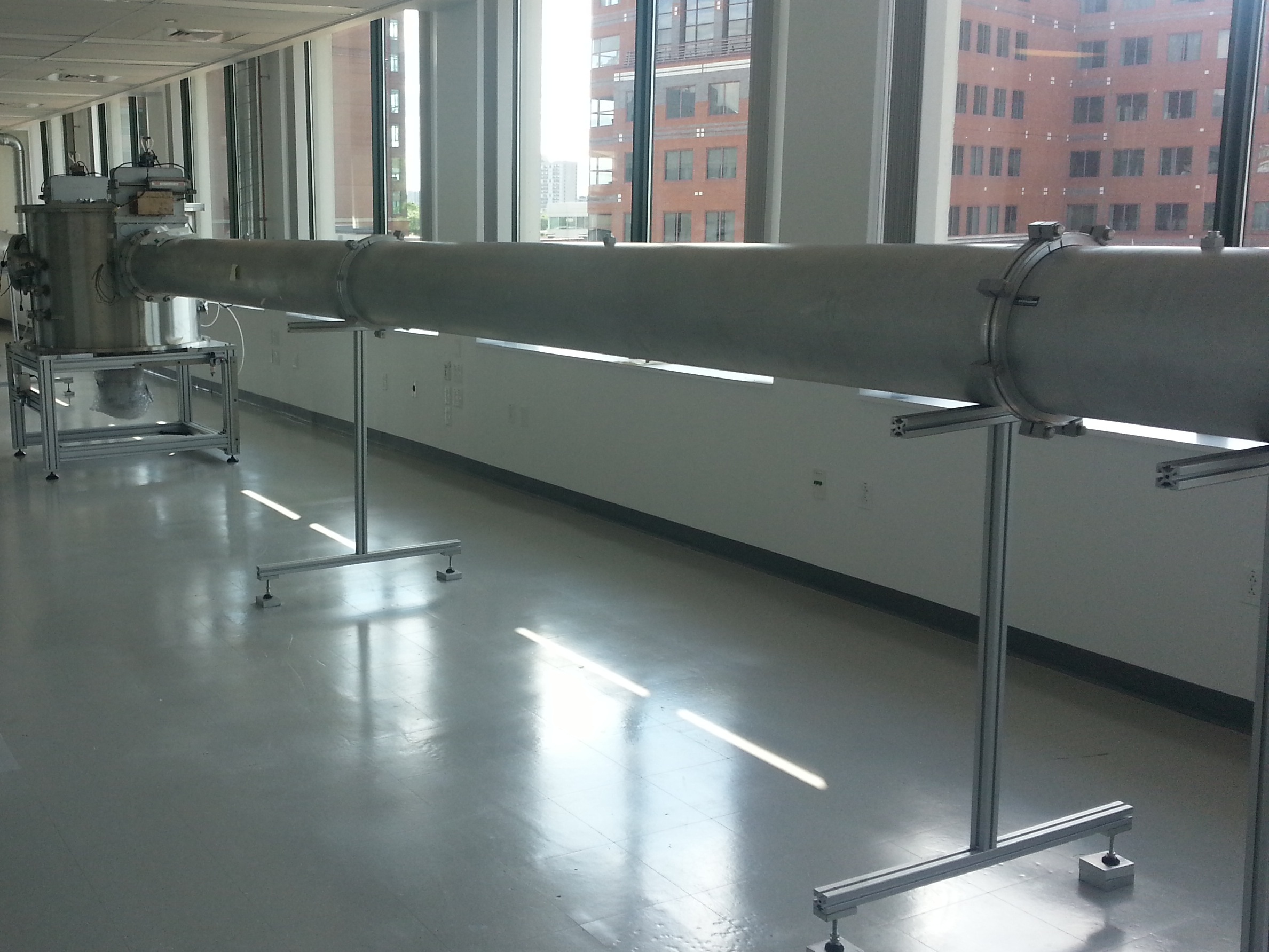}
 \caption{
Polarimetry Beamline, as partially assembled in NE83.
 Shown are the grating chamber and
 the source side beam pipe after initial leveling but before reconnecting wiring
 and gate valve pneumatics.
}
\label{fig:ne83}
\end{figure}

Due to the termination of the lease for the floor occupied by the facility for
the past twenty years, the beamline was moved to a newly leased floor of an adjacent
building, designated NE83.  The beamline was disassembled and is in the
process of reassembly.  Fig.~\ref{fig:ne83} shows the grating chamber and
the source side beam pipe after initial leveling but before reconnecting wiring.
After alignment, the first scientific tests will be to repeat the initial
Phase III tests using the Ti-L and C-K lines.

\section{A Soft X-ray Polarizing Spectrometer}
\label{sec:instr}

\begin{figure}
    \centering
   \includegraphics[height=11cm]{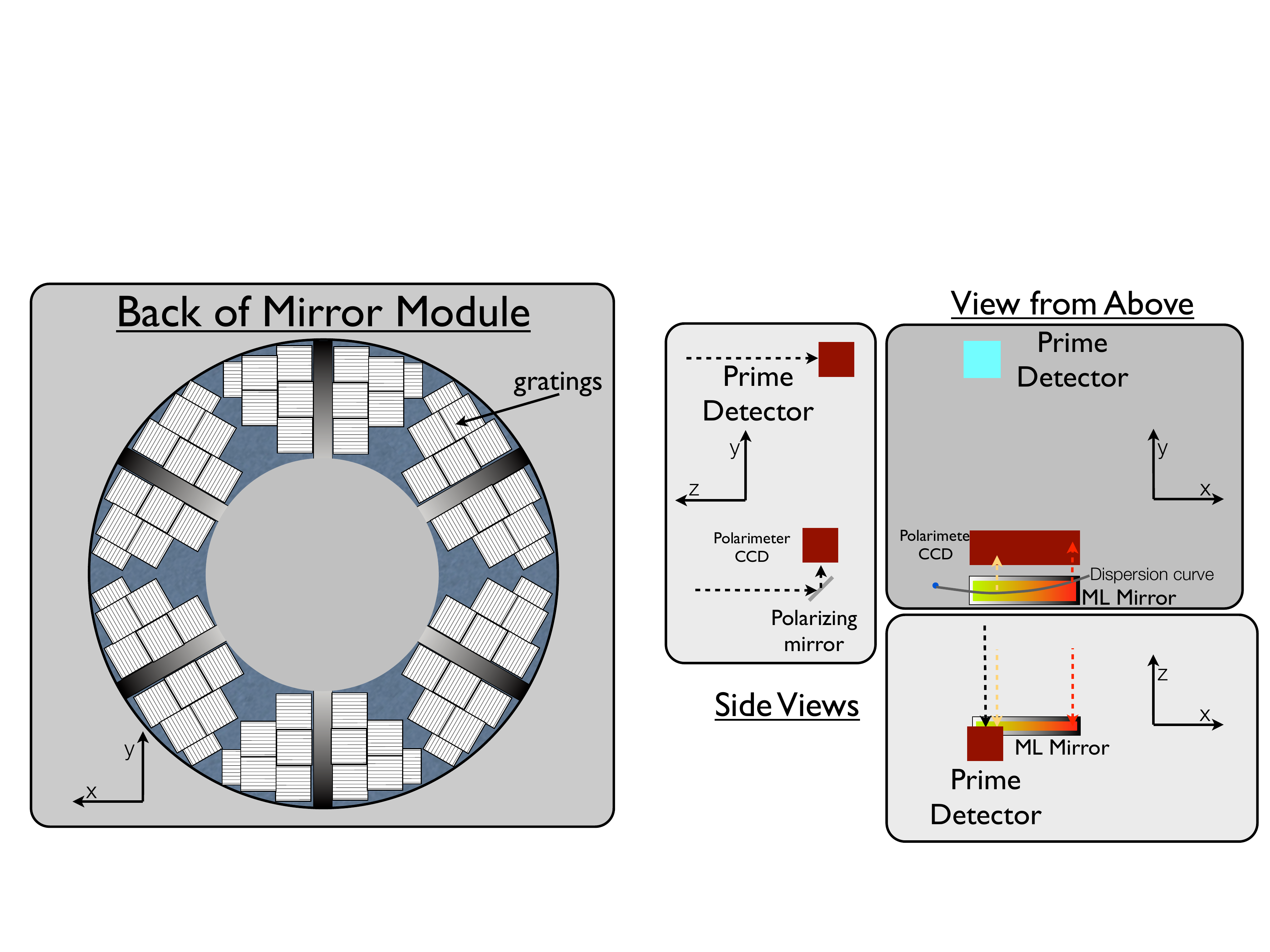}
 \caption{
 Schematic of a suborbital soft X-ray polarimeter using blazed reflection gratings
  in an off-plane configuration\cite{2013ExA....36..389M}.
 {\it Left:} View of the back of the mirror assembly,
    where blazed gratings are oriented approximately azimuthally
    and in sectors.
 {\it Right:} Top and side views of a focal plane layout that could be used for a suborbital rocket
         experiment, in the manner suggested by
	Marshall (2008 \cite{2008SPIE.7011E..63M}).
	The prime detector receives X-rays that do not intercept the grating modules
	The zeroth order is placed at the location of the blue dot so that
	the dispersed spectrum first intercepts the laterally graded multilayer mirror that is angled at
	$45\deg$ to the incoming X-rays.
	In the side view, the dispersion is perpendicular to the plane
	of the drawing and the multilayer mirror is oriented $45\deg$ to the incoming, dispersed X-rays.
}
\label{fig:suborbital}
\end{figure}

The basic design of a polarizing spectrometer
was outlined by Marshall (2008 \cite{2008SPIE.7011E..63M}).  For this paper,
we examine the approach that could be applied to a suborbital rocket experiment.
Figure~\ref{fig:suborbital} shows a possible schematic for a suborbital mission using
blazed gratings such as the Critical Angle Transmission
(CAT) gratings under development at MIT\cite{Heilmann08,2009SPIE.7437E..14H}
or reflection gratings used in an off-plane configuration\cite{2013ExA....36..389M}.
Sampling at least 3 position angles is required in order to measure
three Stokes parameters (I, Q, U) uniquely, so one would require at
least three separate detector
systems (one of which could be just for 0th order)
with accompanying multilayer-coated flats or that the rocket
rotate during the observations (which is expected anyway, to take out
systematic effects).

The system design consists of a mirror system with an assumed effective area of 350 cm$^2$
below 1 keV, backside-illuminated CCD detectors like those on {\em Chandra} with thin
directly deposited optical blocking filters, and CAT or reflection
gratings blazed to maximize efficiency at 300 eV.
We have also computed effective areas using
LETGs \cite{1997SPIE.3113..172P}.
For ML coating reflectivities, we used values that have been achieved in the lab
for single-period MLs used at 45$\deg$ and
interpolated using comparable theoretical models.

The effective area estimate can be used to predict the minimum
detectable polarization (MDP)
for a potential target.
Extragalactic sources such as the BL Lac object
Mk 421 are expected to be highly polarized in the soft X-ray band.
We use the same expected spectrum as assumed in Paper I.
In a 500 s observation of Mk 421,
this instrument could detect polarizations of 3.9\% using CAT or reflection gratings
or 6.5\% using LETGs.
LGMLs with the achievable reflectivities have not yet been
fabricated, so we computed the MDPs for reflectivities as measured by the ALS
for the W/B$_4$C LGML made by RXO.  Interpolating using the Bragg peak reflectivities
at the measured energies gives MDPs of 11.4\% for CAT or reflection gratings, and 16\% using LETGs.
Continued development of LGMLs are
expected to bring these MDPs down below 10\%.

\acknowledgments     %>>>> equivalent to \section*{ACKNOWLEDGMENTS}       

We are very grateful for the assistance and support provided by Steve Kissel
and Beverly LaMarr in providing, operating, and modifying the CCD detector system.
Support for this work was provided by the National Aeronautics and
Space Administration through grant NNX12AH12G and by Research
Investment Grants from the MIT Kavli Institute.

%%%%%%%%%%%%%%%%%%%%%%%%%%%%%%%%%%%%%%%%%%%%%%%%%%%%%%%%%%%%%
%%%%% References %%%%%

\bibliography{polarimetry14}   %>>>> bibliography data in polarimeter10.bib

\begin{thebibliography}{10}

\bibitem{2007SPIE.6688E..31M}
{Marshall}, H.~L., ``{A soft x-ray polarimeter designed for broadband x-ray
  telescopes},'' in [{\em Optics for EUV, X-Ray, and Gamma-Ray Astronomy III.
  Edited by O'Dell, Stephen L.; Pareschi, Giovanni. Proceedings of the SPIE,
  Volume 6688, pp. 66880Z (2007).}{\nolinebreak\hspace{0.1em}]},  {\em
  Presented at the Society of Photo-Optical Instrumentation Engineers (SPIE)
  Conference} {\bf 6688} (Sept. 2007).

\bibitem{2008SPIE.7011E..63M}
{Marshall}, H.~L., ``{Polarimetry with a soft x-ray spectrometer},'' in [{\em
  Society of Photo-Optical Instrumentation Engineers (SPIE) Conference
  Series}{\nolinebreak\hspace{0.1em}]},  {\em Society of Photo-Optical
  Instrumentation Engineers (SPIE) Conference Series} {\bf 7011} (Aug. 2008).

\bibitem{plexas}
{Marshall}, H.~L., {Murray}, S.~S., {Chappell}, J.~H., {Schnopper}, H.~W.,
  {Silver}, E.~H., and {Weisskopf}, M.~C., ``{Realistic, inexpensive, soft
  x-ray polarimeter and the potential scientific return},'' in [{\em
  Polarimetry in Astronomy. Edited by Silvano Fineschi . Proceedings of the
  SPIE, Volume 4843, pp. 360-371 (2003).}{\nolinebreak\hspace{0.1em}]},
  {Fineschi}, S., ed., {\em Presented at the Society of Photo-Optical
  Instrumentation Engineers (SPIE) Conference} {\bf 4843},  360--371 (Feb.
  2003).

\bibitem{2010SPIE.7732E..12M}
{Marshall}, H.~L., {Heilmann}, R.~K., {Schulz}, N.~S., and {Murphy}, K.~D.,
  ``{Broadband soft x-ray polarimetry},'' in [{\em Society of Photo-Optical
  Instrumentation Engineers (SPIE) Conference
  Series}{\nolinebreak\hspace{0.1em}]},  {\em Society of Photo-Optical
  Instrumentation Engineers (SPIE) Conference Series} {\bf 7732} (July 2010).

\bibitem{1994SPIE.2280..257D}
{Dewey}, D., {Humphries}, D.~N., {McLean}, G.~Y., and {Moschella}, D.~A.,
  ``{Laboratory calibration of x-ray transmission diffraction gratings},'' in
  [{\em Proc. SPIE Vol. 2280, p. 257-271, EUV, X-Ray, and Gamma-Ray
  Instrumentation for Astronomy V, Oswald H. Siegmund; John V. Vallerga;
  Eds.}{\nolinebreak\hspace{0.1em}]},  {Siegmund}, O.~H. and {Vallerga}, J.~V.,
  eds., {\em Presented at the Society of Photo-Optical Instrumentation
  Engineers (SPIE) Conference} {\bf 2280},  257--271 (Sept. 1994).

\bibitem{2013SPIE.8861E..1DM}
{Marshall}, H.~L., {Schulz}, N.~S., {Remlinger}, B., {Gentry}, E.~S., {Windt},
  D.~L., and {Gullikson}, E.~M., ``{Progress toward a soft x-ray
  polarimeter},'' in [{\em Society of Photo-Optical Instrumentation Engineers
  (SPIE) Conference Series}{\nolinebreak\hspace{0.1em}]},  {\em Society of
  Photo-Optical Instrumentation Engineers (SPIE) Conference Series} {\bf 8861}
  (Sept. 2013).

\bibitem{2010SPIE.7732E..108M}
{Murphy}, K.~D., {Marshall}, H.~L., {Schulz}, N.~S., {Jenks}, K.~P., {Sommer},
  S.~J.~B., and {Marshall}, E.~A., ``{Soft x-ray polarimeter laboratory
  tests},'' in [{\em Society of Photo-Optical Instrumentation Engineers (SPIE)
  Conference Series}{\nolinebreak\hspace{0.1em}]},  {\em Society of
  Photo-Optical Instrumentation Engineers (SPIE) Conference Series} {\bf 7732}
  (June 2010).

\bibitem{1997SPIE.3113..172P}
{Predehl}, P., {Braeuninger}, H.~W., {Brinkman}, A.~C., {Dewey}, D., {Drake},
  J.~J., {Flanagan}, K.~A., {Gunsing}, T., {Hartner}, G.~D., {Juda}, J.~Z.,
  {Juda}, M., {Kaastra}, J.~S., {Marshall}, H.~L., and {Swartz}, D.~A.,
  ``{X-ray calibration of the AXAF Low Energy Transmission Grating
  Spectrometer: effective area},'' in [{\em Society of Photo-Optical
  Instrumentation Engineers (SPIE) Conference
  Series}{\nolinebreak\hspace{0.1em}]},  {Hoover}, R.~B. and {Walker}, A.~B.,
  eds., {\em Society of Photo-Optical Instrumentation Engineers (SPIE)
  Conference Series} {\bf 3113},  172--180 (July 1997).

\bibitem{Heilmann08}
Heilmann, R.~K., Ahn, M., Gullikson, E.~M., and Schattenburg, M.~L., ``Blazed
  high-efficiency x-ray diffraction via transmission through arrays of
  nanometer-scale mirrors,'' {\em Opt. Express}~{\bf 16}(12),  8658--8669
  (2008).

\bibitem{2009SPIE.7437E..14H}
{Heilmann}, R.~K., {Ahn}, M., {Bautz}, M.~W., {Foster}, R., {Huenemoerder},
  D.~P., {Marshall}, H.~L., {Mukherjee}, P., {Schattenburg}, M.~L., {Schulz},
  N.~S., and {Smith}, M., ``{Development of a critical-angle transmission
  grating spectrometer for the International X-Ray Observatory},'' in [{\em
  Society of Photo-Optical Instrumentation Engineers (SPIE) Conference
  Series}{\nolinebreak\hspace{0.1em}]},  {\em Society of Photo-Optical
  Instrumentation Engineers (SPIE) Conference Series} {\bf 7437} (Aug. 2009).

\bibitem{2013ExA....36..389M}
{McEntaffer}, R., {DeRoo}, C., {Schultz}, T., {Gantner}, B., {Tutt}, J.,
  {Holland}, A., {O'Dell}, S., {Gaskin}, J., {Kolodziejczak}, J., {Zhang},
  W.~W., {Chan}, K.-W., {Biskach}, M., {McClelland}, R., {Iazikov}, D., {Wang},
  X., and {Koecher}, L., ``{First results from a next-generation off-plane
  X-ray diffraction grating},'' {\em Experimental Astronomy}~{\bf 36},
  389--405 (Aug. 2013).

\end{thebibliography}
\bibliographystyle{spiebib}   %>>>> makes bibtex use spiebib.bst

\end{document}